\documentclass[aps,prb,twocolumn,showpacs,dvipdfm,superscriptaddress,floatfix]{revtex4}

\usepackage{graphicx,amssymb,amsfonts,amsmath}
\usepackage[dvipdfm]{hyperref}
\hypersetup{colorlinks=true,linkcolor=blue,anchorcolor=blue,citecolor=blue,filecolor=blue,
urlcolor=blue,bookmarksnumbered=true,pdfview=FitB} %

\def\rn{{\rho_\text{n}}}

\def\rs{{\rho_\text{s}}}
\def\rD{{\rho_\Delta}}

\def\bvs{{\bf v}_\text{s}}
\def\bvn{{\bf v}_\text{n}}

\def\vs{{v_\text{s}}}
\def\vn{{v_\text{n}}}

\def\mmu{\tilde{\mu}}
\def\llam{\tilde{\lambda}}

\def\half{\frac{1}{2}}
\def\lag{\mathcal{L}}
\def\d{\delta}
\def\p{\partial}
\def\nab{\mbox{\boldmath $\nabla$}}

\begin{document}

\title{Hydrodynamic theory of supersolids: Variational principle and effective Lagrangian}

\author{C.-D. Yoo}

\author{Alan T. Dorsey}

\affiliation{Department of Physics, University of Florida, P.O. Box
118440, Gainesville, FL 32611-8440}

\date{\today}

\begin{abstract}

We develop an effective low-energy, long-wavelength theory of a bulk
supersolid---a putative phase of matter with simultaneous
crystallinity and Bose condensation. Using conservation laws and
general symmetry arguments we derive an effective action that
correctly describes the coupling between the Bose condensation and
the elasticity of the solid. We use our effective action to
calculate the correlation and response functions for the supersolid,
and we show that the onset of supersolidity produces peaks in the
response function, corresponding to propagating second sound modes
in the solid. Throughout our work we make connections to existing
work on effective theories of superfluids and normal solids, and we
underscore the importance of conservation laws and symmetries in
determining the number and character of the collective modes.

\end{abstract}


\maketitle

\section{Introduction}

In 1969 Andreev and Lifshitz proposed a novel phase of matter in
quantum Bose crystals, wherein a Bose condensate of point defects
would coexist with the crystallinity of the solid.\cite{andreev69}
This is perhaps the most conceptually clear picture of what we now
call a ``supersolid,'' although suggestions of the coexistence (or
non-coexistence) of Bose condensation and crystallinity can be
traced to the earlier work of Penrose and Onsager \cite{penrose56}
and Chester.\cite{chester70} Andreev and Lifshitz provided an
elegant (albeit incomplete) formulation of the hydrodynamics of
supersolids, and predicted propagating modes analogous to second (or
fourth) sound in liquid $^4$He. Their hydrodynamic formulation was
further extended  by Saslow \cite{saslow77} and by Liu.\cite{liu78}
Experimental searches for signatures of the supersolid phase proved
fruitless\cite{meisel92} until recently, when Kim and Chan observed
rotational inertia anomalies in solid $^4$He that they interpreted
as evidence for supersolidity.\cite{kim04.1, kim04.2, kim05}  Their
work fueled extensive searches for further evidence of this elusive
phase of matter, \cite{rittner06,day06,diallo07,rittner07,
kondo07,penzev08,blackburn07,aoki07} and there are now a number of
extensive reviews of the experimental and theoretical progress in
this area---see Refs.~\onlinecite{prokofev07},
\onlinecite{balibar08}, \onlinecite{chan08}, and
\onlinecite{galli08}.

The present work is a detailed---and we believe novel---study of the
hydrodynamics of bulk supersolids, of the type originally proposed
by Andreev and Lifshitz. Our work uses conservation laws and general
symmetry principles to derive an effective action for a supersolid.
We rely extensively on a variational principle used to obtain the
dynamic equations in various continuum systems: normal fluids,
\cite{lin63, seliger68, salmon88} superfluids,\cite{zilsel50, lin63,
jackson78, geurst80, purcell81, coste98} normal solids,
\cite{seliger68, holm83} liquid crystals,\cite{holm01} trapped
superfluid gases,\cite{taylor05} and relativistic fluids.
\cite{jackiw04} This effective action is a powerful tool for
calculating and elucidating the collective modes of the supersolid
phase, and one of our important new results is a calculation of the
correlation and response functions in the supersolid phase. Moving
beyond linearized hydrodynamics, the effective action can also be
use to study the dynamics and interaction of topological
defects---vortices and dislocations---in the supersolid, the topics
of subsequent publications.\cite{yoo08.3} We should also state what
this work is \emph{not}---it is not an explanation of the recent
experiments on possible supersolidity in $^4$He, as the prevailing
wisdom suggests that structural disorder plays a key role in most of
the experiments, and our simplified model assumes an ordered solid.

This work is organized as follows. In Sec.~II we derive the
supersolid hydrodynamics and the effective Lagrangian density using
the variational principle. We show that the equation of motion are
equivalent (up to a term nonlinear in the elastic strain) to the
hydrodynamic equations of motion derived by Andreev and Lifshitz
\cite{andreev69}. We also discuss the connection to the work of
Saslow,\cite{saslow77} Liu,\cite{liu78} and Son.\cite{son05} In
Sec.~III we use a quadratic version of the Lagrangian to investigate
the linearized hydrodynamics of a supersolid. Finally, in Sec.~IV
the collective modes and the density-density correlation function of
a model supersolid are calculated in detail. The Appendices provide
additional detail, as an aid to the reader.

\section{Variational Principle and an Effective Lagrangian of Supersolids}

We start with a Lagrangian density for a supersolid in the Eulerian
description (in which all quantities are depicted at fixed position
$\mathbf{x}$ and time $t$),
\begin{eqnarray}
\lag_\text{SS} &=& \half \rs_{ij} \vs_i \vs_j
+ \half (\rho \d_{ij} -\rs_{ij}) \vn_i \vn_j
\nonumber
\\
&&
- U_\text{SS}(\rho,\rs_{ij},s,R_{ij}),
\label{lagrangian1}
\end{eqnarray}
where $\rs_{ij}$ is the superfluid density tensor, $\rho$ is the
total density, $\bvs$ is the velocity of the super-components,
$\bvn$ is the velocity of the normal-components, $s$ is the entropy
density, and
\begin{equation}
R_{ij} \equiv \p_i R_j
\end{equation}
is the deformation tensor, with ${\bf R}$ the coordinate affixed to
material elements ($\partial_i\equiv
\partial/\partial x_i$ and $\partial_t \equiv \partial/\partial_t$ in what follows).
The first two terms in the Lagrangian density are the kinetic energy
densities of the super-component and the normal-component,
respectively, and the third term is the internal energy density
which is a function of $\rho$, $\rs_{ij}$, $s$, and $R_{ij}$. In
contrast to a superfluid, $R_{ij}$ appears explicitly in
$U_\text{SS}$ for a supersolid, a reflection of the solid's broken
translational symmetry. As shown in Appendix~B, the internal energy
density satisfies the thermodynamic relation
\begin{eqnarray}
d U_\text{SS} &=&
T ds + \left[ \mu + \half (\vn_i - \vs_i)^2 \right] d \rho
- \lambda_{ik} d R_{ik}
\nonumber
\\
&&
- \half (\vn_i - \vs_i)(\vn_j - \vs_j) d{\rho_s}_{ij},
\label{thermo.Uss}
\end{eqnarray}
where $\mu$ is the chemical potential per unit mass, and
$\lambda_{ij}$ the stress tensor. Given the Lagrangian density in
Eq.~(\ref{lagrangian1}), the action is
\begin{equation}
S_\text{SS} = \int dt\int d^3x \lag_\text{SS} .
\end{equation}

The equations of motion for a supersolid are obtained from
variations of $S_\text{SS}$ with respect to the dynamical variables.
However, as illustrated in Appendix~A, the dynamical variables are
not independent and one must insure that conservation laws and
broken symmetries are incorporated in the action through the use of
auxiliary fields (Lagrange multipliers). For a three dimensional
supersolid there are five conserved quantities: the mass, the
entropy and the three components of the momentum. Among these
constraints we impose only the mass and entropy conservation laws,
and show below that the momentum conservation is the byproduct of
the variational principle. Conservation of mass is expressed through
the equation of continuity,
\begin{equation}
\p_t \rho + \p_i j_i=0, \label{SS.continuity}
\end{equation}
where the mass current $j_i$ is
\begin{equation}
j_i =\rs_{ij} \vs_j + (\rho \d_{ij} - \rs_{ij}) \vn_j.
\label{mass.current}
\end{equation}
The entropy conservation law is
\begin{equation}
\p_t s + \p_i (s \vn_i) = 0,
\label{SS.entropy}
\end{equation}
in which only $\bvn$ is involved because the entropy is transported
by the normal component. Finally, we account for the broken
translational symmetry using \emph{Lin's constraint}, \cite{lin63}
\begin{equation}
\frac{D_n R_i}{D t} =0,
\label{Lin}
\end{equation}
where $D_n / D t \equiv \p_t + \vn_i \p_i$. This constraint states
that the Lagrangian coordinates (i.e., the initial positions of
particles) do not change along the paths of the normal component.
Indeed, Lin's constraint was first introduced to generate vorticity
in the Lagrangian description of an isentropic normal fluid.
\cite{lin63} We incorporate all of the constraints,
Eqs.~(\ref{SS.continuity})-(\ref{Lin}), into the Lagrangian density
Eq.~(\ref{lagrangian1}) by using the Lagrange multipliers $\alpha$,
$\phi$, and $\beta_i$, with the result:
\begin{eqnarray}
\lag_\text{SS} &=&
\half \rs_{ij} \vs_i \vs_j + \half (\rho \d_{ij} -\rs_{ij})\vn_i \vn_j
\nonumber
\\
&&
- U_\text{SS}(\rho,\rs_{ij},s,R_{ij})
+ \alpha \bigg[ \p_t s + \p_i (s \vn_i) \bigg]
\nonumber
\\
&&
+ \phi \Bigg\{\p_t \rho + \p_i \bigg[ \rs_{ij} \vs_j + (\rho \d_{ij} - \rs_{ij})
\vn_j \bigg]\Bigg\}
\nonumber
\\
&&
+ \beta_i \bigg[ \p_t(sR_i) + \p_j (s R_i \vn_j) \bigg].
\label{lag2}
\end{eqnarray}
Note that in our formulation  Lin's constraint is combined with the
entropy conservation law.

We are now in a position to derive the hydrodynamic equations of
motion for supersolids. First of all, the variation of the action
with respect to $\vs_i$ produces
\begin{equation}
\vs_i = \p_i \phi.
\label{SS.velocity}
\end{equation}
Therefore, the superfluid component of the velocity is a potential
flow, as expected [rotational flow can be obtained by introducing
another constraint; see Ref.~\onlinecite{geurst80}]. The remaining
equations of motion are
\begin{itemize}
\item $\d \rho:$
\begin{equation}
\half v_n^2 - \frac{\p U_\text{SS}}{\p \rho} -\p_t \phi - \vn_i \vs_i = 0,
\label{eqnphi}
\end{equation}
\item $\d \rs_{ij}:$
\begin{equation}
\frac{\p U_\text{SS}}{\p \rs_{ij}} = - \half (\vs_i - \vn_i) (\vs_j
- \vn_j), \label{SSeqnrhos}
\end{equation}
\item $\d s$:
\begin{equation}
\frac{D_n \alpha}{D t}
+ R_i \frac{D_n \beta_i}{D t}
+ \frac{\p U_\text{SS}}{\p s} = 0,
\label{SSeqnalpha1}
\end{equation}
\item $\d \vn_i$:
\begin{equation}
\p_i \alpha + R_j \p_i \beta_j =
\frac{1}{s} (\rho \d_{ij} - \rs_{ij})(\vn_j - \vs_j),
\label{SSeqnalpha2}
\end{equation}
\item $\d R_i$:
\begin{equation}
\frac{D_n \beta_i}{D t}
- \frac{1}{s}\p_j \left(\frac{\p U_\text{SS}}{\p R_{ji}}\right) = 0.
\label{SS.eqn.beta}
\end{equation}
\end{itemize}
In the above equations of motion we have eliminated the gradient of
$\phi$ by using Eq.~(\ref{SS.velocity}). In addition to the derived
equations of motion, the variations with respect to the Lagrange
multipliers reproduce the imposed constraints,
Eqs.~(\ref{SS.continuity})-(\ref{Lin}). Therefore,
Eqs.~(\ref{SS.continuity})-(\ref{Lin}),
(\ref{SS.velocity})-(\ref{SS.eqn.beta}) are the hydrodynamic
equations for supersolids.

In the following we demonstrate that the equations of motion derived
above are equivalent to the non-dissipative supersolid hydrodynamics
developed by Andreev and Lifshitz, \cite{andreev69} Saslow,
\cite{saslow77} and Liu. \cite{liu78} First, the taking the gradient
of Eq.~(\ref{eqnphi}) produces the Josephson equation
\begin{equation}
\p_t \vs_i = - \p_i \mu - \half \p_i \vs^2,
\label{josephson.eqn}
\end{equation}
where we have used the thermodynamic relation for $\p U_\text{ss} /
\p \rho$ given in Eq.~(\ref{thermo.Uss}). Second, we derive the
momentum conservation equation; the following identity simplifies
the derivation:
\begin{equation}
\frac{D (a \p_i b)}{D t} = \p_i b \frac{D a}{D t} + a \p_i \left(
\frac{D b}{D t} \right) - a \p_j b \p_i v_j,
\end{equation}
where $D / D t \equiv \p_t + v_i \p_i$. Take $D_n / D t$ of
Eq.~(\ref{SSeqnalpha2}), and eliminate the Lagrange multipliers by
using Eqs.~(\ref{SS.entropy}), (\ref{Lin}),
(\ref{SSeqnalpha1})-(\ref{SS.eqn.beta}). The result is
\begin{equation}
\begin{split}
& \hspace{0.5cm}
\frac{D_n }{D t} \bigg[ (\rho \d_{ij} - \rs_{ij}) (\vn_j - \vs_j) \bigg] =
- s \p_i \left( \frac{\p U_\text{SS}}{\p s} \right)
\\
&
- \p_i R_j \p_k \left( \frac{\p U_\text{SS}}{\p R_{kj}} \right)
- (\rho \d_{ij} - \rs_{ij}) (\vn_j - \vs_j) \p_k \vn_k
\\
& \hspace{1.5cm}
-(\rho \d_{jk} - \rs_{jk}) (\vn_k - \vs_k) \p_i \vn_j.
\label{eqn1}
\end{split}
\end{equation}
Third, combine Eq.~(\ref{eqn1}) with the thermodynamic relation,
Eq.~(\ref{thermo.Uss}), the continuity equation,
Eq.~(\ref{SS.continuity}), and the Josephson equation,
Eq.~(\ref{josephson.eqn}). After some algebra, we obtain the
momentum conservation law
\begin{equation}
\partial_t j_i + \partial_j \Pi_{ij} = 0,
\label{momentum.conservation.eqn}
\end{equation}
where $j_i$ is the mass current given in Eq.~(\ref{mass.current}),
and $\Pi_{ij}$ is the (non-dissipative) stress tensor
\begin{eqnarray}
\Pi_{ij} &=& \rho\vs_i\vs_j + \vs_i p_j + \vn_j p_i
- R_{ik}\lambda_{jk}
\nonumber
\\
&&
-\bigg[\epsilon -
Ts - \mu \rho - (\vn_j - \vs_j) p_j \bigg]\delta_{ij}, 
\label{stress.tensor}
\end{eqnarray}
where $p_i \equiv (\rho \d_{ij} - \rs_{ij}) (\vn_j - \vs_j)$ and
$\epsilon$ satisfies a thermodynamic relation given by
Eq.~(\ref{SS.energy.1}). Note that the Josephson equation,
Eq.~(\ref{josephson.eqn}), and the momentum conservation equation,
Eq.~(\ref{momentum.conservation.eqn}), are Eqs.~(9) and (12) of
Andreev and Lifshitz~\cite{andreev69} [Andreev and Lifshitz
neglected nonlinear strain terms, effectively replacing $R_{ik}$ by
$\delta_{ik}$ in the last term of Eq.~(\ref{stress.tensor}) above].
Moreover, the momentum conservation equation is equivalent to
Eq.~(4.16) of Saslow~\cite{saslow77} when $\bvs$ is taken as a
Galilean velocity, and Eq.~(3.40) of Liu~\cite{liu78} in the case
where the superthermal current vanishes.

The Lagrangian density used to derive the hydrodynamics of
supersolids, Eq.~(\ref{lag2}), can be recast into a more familiar
and compact form by using the equations of motion, as illustrated
for an ideal fluid in Appendix~A. To see this, we integrate the
terms involving the Lagrange multipliers by parts (neglecting
boundary terms), and use Eqs.~(\ref{SS.velocity}) and
(\ref{SSeqnalpha1}) to eliminate $\alpha$ and $\beta_i$. We then
obtain
\begin{eqnarray}
\lag_\text{SS} &=&
-\rho \p_t \phi - \half \rs_{ij} \p_i \phi \p_j \phi
+ \half (\rho \d_{ij} -\rs_{ij})\vn_i \vn_j
\nonumber
\\
&&
- (\rho \d_{ij} -\rs_{ij}) \vn_j \p_i \phi
- f(\rho,\rs_{ij},T,R_{ij}),
\label{lag4}
\end{eqnarray}
where $f \equiv U_\text{SS} - Ts$ satisfies the thermodynamic
relation
\begin{eqnarray}
d f &=&
- s dT + \left[ \mu + \half (\vn_i - \vs_i)^2 \right] d \rho
- \lambda_{ik} d R_{ik}
\nonumber
\\
&&
- \half (\vn_i - \vs_i)(\vn_j - \vs_j) d\rs_{ij}.
\label{thermodynamic.relation.f}
\end{eqnarray}
When cast in this form, we see that the coupling between the
superfluid and the normal fluid [the fourth term in
Eq.~(\ref{lag4})] is $- (\rho \d_{ij} -\rs_{ij})\vn_i\vs_j$; this is
a ``current-current'' interaction, where the coupling constant is
the normal fluid density. This coupling coefficient is
\emph{universal}--it is determined by conservation laws and Galilean
invariance.

As mentioned earlier, several other authors have recently proposed
Lagrangian descriptions for supersolids. Son \cite{son05} used
symmetry-based arguments to derive an effective Lagrangian for a
supersolid. To connect to Son's results, we first invert Lin's
constraint, Eq.~(\ref{Lin}), to obtain
\begin{equation}
{v_n}_i = - R_{ji}^{-1} \p_t R_j, \label{inverted.Lin}
\end{equation}
where $R_{ji}^{-1} \equiv \p x_i / \p R_j$ and $R_{ij} R_{jk}^{-1} =
\d_{ik}$. Substituting this into Eq.~(\ref{lag4}), we obtain a
Lagrangian similar in form to Eq.~(23) of Son's paper (however, our
energy density $f$ depends upon $\rho$, $\rho_s$, and $T$ in
addition to $R_{ij}$). A different approach was used by Josserand
\emph{et al.}, \cite{josserand07} who applied a homogenization
procedure to a nonlocal version of the Gross-Pitaevskii equation to
systematically derive a long-wavelength Lagrangian for a supersolid.
On the whole, our Eq.~(\ref{lag4}) agrees with their Eq.~(4), once
we identify their $\rho(n)$ with our normal fluid density
$\rho\delta_{ij} - \rs_{ij}$.  Finally, Ye \cite{ye08} proposed a
phenomenological supersolid Lagrangian; however, the Lagrangian in
his Eq.~(1) is not manifestly Galilean invariant, so that his
coupling constants $a_{\alpha\beta}$ are arbitrary. His approach
also misses certain nonlinear terms that are important when treating
topological defects.

\section{Quadratic Lagrangian Density and the Linearized Hydrodynamics of Supersolids}

In this section we discuss the propagation of collective modes in
supersolids by examining the response to small fluctuations away
from equilibrium. The number of collective hydrodynamic modes of a
system can be inferred by enumerating the system's conservation laws
and broken symmetries.\cite{martin72} Since we are more interested
in the effect of density or defect fluctuations than thermal
fluctuations, for simplicity we ignore thermal fluctuations in what
follows. For a three dimensional supersolid there are conservation
laws for mass, three components of momentum, and energy; however, by
ignoring thermal fluctuations we can omit the energy conservation
law. In addition the conservation laws, there are three broken
translational symmetries (due to the crystallinity) and one broken
gauge symmetry (due to the Bose-Einstein condensation). Thus, a
three dimensional supersolid without thermal fluctuations has eight
conservation laws and broken symmetries (nine, if conservation of
energy is included). Correspondingly, there are eight hydrodynamic
modes: two pairs of the ordinary transverse propagating modes, a
pair of longitudinal first sound modes, and a pair of longitudinal
second sound modes (note that a solution of the hydrodynamic
equations with dispersion $\omega = \pm ck$ would count as
\emph{two} modes--a pair of modes, with one propagating and a second
counter-propagating). The appearance of the longitudinal second
sound modes is one of the key signatures of a supersolid.

We start by establishing the notation for the small fluctuations
away from equilibrium. The equilibrium value of the densities will
be denoted with a subscript of `0', and the density fluctuations
will be denoted by $\d\rho$, so that $\rho = \rho_0 + \d \rho$ and
$\rs_{ij} = {{\rs}_0}_{ij} + \d \rs_{ij}$. For the lattice
fluctuations, let ${\bf u}$ denote the (small) deformation field
away from the unstrained solid (i.e. the difference between the
Lagrangian and the Eulerian coordinates):
\begin{equation}
{\bf x} = {\bf R} + {\bf u}.
\end{equation}
Then the deformation tensor becomes
\begin{equation}
R_{ij} = \d_{ij} - w_{ij},
\end{equation}
where $w_{ij} \equiv \p_i u_j$. Finally, the velocity of the normal
component is obtained by linearizing the inverted Lin's constraint,
Eq.~(\ref{inverted.Lin}), so that to lowest order in the strain
field the normal velocity is the time derivative of the displacement
field,
\begin{equation}
\vn_i = \p_t u_i.
\end{equation}
Expanding the Lagrangian density, Eq.~(\ref{lag4}), up to second
order in the small quantities $\d \rho$, $\d \rs_{ij}$, $w_{ij}$,
$\p_i \phi$ and $\p_t \phi$, we obtain
\begin{eqnarray}
\lag_\text{SS}^\text{quad} &=&
- \rho_0 \p_t \phi - {\lambda_0}_{ij} w_{ij}
- \mu_0 \d \rho
- \d \rho \p_t \phi
\nonumber
\\
&& - \half \rho_0 (\p_i \phi)^2 - \frac{\p \mu}{\p
w_{ij}}\bigg|_\rho \d \rho \, w_{ij} - \half \frac{\p \mu}{\p
\rho}\bigg|_{w_{ij}} (\d \rho)^2 \nonumber
\\
&&
+ \half {\rn_0}_{ij} \left( \p_t u_i - \p_i \phi \right)
\left( \p_t u_j - \p_j \phi \right)
\nonumber
\\
&&
- \half \frac{\p \lambda_{ij}}{\p w_{lk}}\bigg|_\rho w_{ij} w_{lk},
\label{SS.quad.lag}
\end{eqnarray}
where ${\rn_0}_{ij} \equiv {\rho_0}_{ij} - {\rs_0}_{ij}$ and the
thermodynamic relation for $f$,
Eq.~(\ref{thermodynamic.relation.f}), is used. In the above
expansion we have dropped constants which do not contribute to the
equations of motion, and have neglected terms proportional to $\p f
/ \p \rs_{ij}$ because they are of higher order [see
Eq.~(\ref{SSeqnrhos})]; consequently, the quadratic Lagrangian turns
out to be independent of fluctuations of the superfluid density.
However, we have kept the first two terms in
Eq.~(\ref{SS.quad.lag}); although they are total derivatives and
would seem to be irrelevant to the equations of motion, they are
non-trivial for topological defects such as vortices or
dislocations. In fact, one can show \cite{yoo08.3} that the first
term produces the Magnus force on a vortex \cite{zhu96} and the
second term generates the Peach-Koehler force on a dislocation.
\cite{peach50, landau86} We will defer the discussion of these
effects to a subsequent publication. \cite{yoo08.3}

Now we are in a position to study the hydrodynamic modes of a
supersolid. The quadratic Lagrangian density,
Eq.~(\ref{SS.quad.lag}), produces three linearized equations of
motions which are equivalent to the continuity equation, the
Josephson equation and the momentum conservation equation. Before
proceeding further, it is useful to rewrite the quadratic Lagrangian
density in terms of the defect density fluctuation by using one of
the equations of motion. By varying the action with respect to $\d
\rho$ we obtain
\begin{equation}
\d \rho =
\frac{\p \rho}{\p w_{ij}}\bigg|_{\mu} w_{ij}
+ \frac{\p \rho}{\p \mu}\bigg|_{w_{ij}} \d \mu,
\label{quad.rho}
\end{equation}
where we have used the linearized Josephson equation ($\p_t \phi = -
\mu_0 - \d \mu$) and the identity
\begin{equation}
\frac{\p x}{\p y}\bigg|_z =
-\frac{\p z}{\p y}\bigg|_x \frac{\p x}{\p z}\bigg|_y.
\label{identity1}
\end{equation}
From Eq.~(\ref{quad.rho}) it is clear that the density fluctuation
is an independent hydrodynamic variable--it is not slaved to the
lattice deformation, as would be the case for a commensurate solid,
where $\d \rho = - \rho_0 \nab \cdot {\bf u}$. Indeed, in a real
(incommensurate) crystal a density fluctuation can be produced by
lattice deformations or by point defects (vacancies and
interstitials). To highlight the role of defects we will introduce
the defect density fluctuation $\d \rD$ as our independent
hydrodynamic variable, instead of $\d \rho$. The local defect
density is defined as the difference between the density of
vacancies, $\rho_\text{V}$, and the density of interstitials,
$\rho_\text{I}$:
\begin{equation}
\rD = \rho_\text{I} - \rho_\text{V}.
\end{equation}
The minus sign is necessary so that the total defect density is
conserved--in the bulk of the crystal vacancies and interstitials
are created and destroyed in pairs (ignoring surface effects). Then
we have
\begin{equation}
\d \mu = \frac{\p \mu}{\p w_{ij}}\bigg|_{\rD} w_{ij}
+ \frac{\p \mu}{\p \rD}\bigg|_{w_{ij}} \d \rD,
\label{thermo.identity.mu}
\end{equation}
and from Eq.~(\ref{quad.rho}) we obtain
\begin{equation}
\d \rho = \frac{\p \rho}{\p w_{ij}}\bigg|_{\rD} w_{ij} +\frac{\p
\rho}{\p \rD}\bigg|_{w_{ij}} \d \rD, \label{quadrho2}
\end{equation}
where we have used Eq.~(\ref{identity1}) and the identity
\begin{equation}
\frac{\p x}{\p y}\bigg|_0 =
\frac{\p x}{\p y}\bigg|_z + \frac{\p x}{\p z}\bigg|_y\frac{\p z}{\p y}\bigg|_0.
\end{equation}
Equation~(\ref{quadrho2}) shows that a density fluctuation in an
isothermal supersolid is caused either by a lattice deformation or
by a defect density fluctuation $\d \rD$, just as in a normal solid.
\cite{zippelius80, ostlund82, chaikin95} Following
Zippelius~\textit{et al.} (ZHN), \cite{zippelius80} we can identify
$(\p \rho / \p w_{ij})_{\rD} = -\rho_0 \d_{ij}$ and $(\p \rho / \p
\rD)_{w_{ij}} = 1$. We finally obtain
\begin{equation}
\d \rho = -\rho_0 w_{ii} + \delta\rho_\Delta, \label{quadrho1}
\end{equation}
which illustrates the roles of the lattice deformation
$w_{ii}=\nabla\cdot {\bf u}$ and the defect density fluctuation in
determining the total density fluctuation. We note in passing that
in a higher order expansion of the Lagrangian density the terms
proportional to the superfluid density fluctuation must also be
retained in Eq.~(\ref{quadrho1}). This would resemble the
``three-fluid'' scenario proposed by Saslow \cite{saslow05} in which
the lattice density and velocity are introduced for the third fluid
component.

We can now use Eq.~(\ref{quadrho1}) to rewrite the quadratic
Lagrangian density in terms of the defect density, with the result
\begin{eqnarray}
\lag_\text{SS}^\text{quad} &=&
\rho_0 w_{ii} \p_t \theta
- {\rn_0}_{ij} \p_t u_i \p_j \theta
- \half \rho_0^2 \frac{\p \mu}{\p \rho}\bigg|_{w_{ij}} w_{ii}^2
\nonumber
\\
&&
- \d \rD \p_t \theta
- \half {\rs_0}_{ij} \p_i \theta \p_j \theta
- \frac{\p \mu}{\p w_{ij}}\bigg|_\rho \d \rD w_{ij}
\nonumber
\\
&&
+ \rho_0 \frac{\p \mu}{\p \rD}\bigg|_{w_{ij}} w_{ii} \d \rD
- \half \frac{\p \mu}{\p \rD}\bigg|_{w_{ij}} \d \rD^2
\nonumber
\\
&&
- \half \frac{\p \lambda_{ji}}{\p w_{lk}}\bigg|_\rho w_{ij}w_{lk}
+ \rho_0 \frac{\p \mu}{\p w_{ij}}\bigg|_\rho w_{ij}w_{kk}
\nonumber
\\
&&
+ \half {\rn_0}_{ij} \p_t u_i \p_t u_j,
\label{SS.quad.lag.defect}
\end{eqnarray}
where we introduced $\theta = \phi + \mu_0 t$. Next, we derive the
linearized equations of motion from the Lagrangian density. First,
note that the variation with respect to $\d \rD$ reproduces
Eq.~(\ref{thermo.identity.mu}) because $\p_t \theta$ is $-\d \mu$.
Second, taking the variation with respect to $\theta$ produces the
linearized equation of continuity, expressed in terms of $\d \rD$:
\begin{equation}
\p_t \d \rD + \partial_i j^\Delta_i = 0, \label{defect.density}
\end{equation}
where the defect current density is given by
\begin{equation}
j^\Delta_i = \rho_{s0ij}(\partial_j \theta - \partial_t u_j).
\end{equation}
We see that the defect current arises from counterflow between the
superfluid velocity $\nabla\theta$ and the normal fluid velocity
$\partial_t {\bf u}$, and vanishes when $\rho_{s0ij}=0$, in the
normal state. In other words, $\p_t \d \rD = 0$ in the normal state,
in agreement with the non-dissipative description of normal
solids\cite{zippelius80} in which defect currents are only produced
through diffusion (i.e., the defect current is dissipative in the
normal solid). The last equation of motion derived from the
variation of $u_i$ is
\begin{equation}
\begin{split}
&
{\rn_0}_{ij} \p_t^2 u_j
- \left( \frac{\p \mu}{\p w_{ji}}\bigg|_\rho
- {\rn_0}_{ij} \frac{\p \mu}{\p \rD}\bigg|_{w_{ij}} \right) \p_j \d \rD
\\
& \hspace{0.1cm}
-\left( \frac{\p \lambda_{ji}}{\p w_{lk}}\bigg|_\rho
- {\rn_0}_{ij} \frac{\p \mu}{\p w_{lk}}\bigg|_\rD
- \rho_0 \frac{\p \mu}{\p w_{ij}}\bigg|_\rho \d_{lk} \right)  \p_j w_{lk} = 0.
\label{eqn.motion.u}
\end{split}
\end{equation}
When the time derivative of Eq.~(\ref{defect.density}) is combined
with Eq.~(\ref{thermo.identity.mu}), we obtain
\begin{equation}
\begin{split}
&\p_t^2 \d \rD
- {\rs_0}_{ij} \frac{\p \mu}{\p \rD}\bigg|_{w_{ij}} \p_i \p_j \d \rD
\\
& \hspace{1cm} - {\rs_0}_{ij} \p_i \p_t^2 u_j - {\rs_0}_{ij}
\frac{\p \mu}{\p w_{lk}}\bigg|_{\rD} \p_i \p_j w_{lk} = 0.
\label{eqn.motion.defect}
\end{split}
\end{equation}
Our linearized equations of motion, Eqs.~(\ref{eqn.motion.u}) and
(\ref{eqn.motion.defect}), are equivalent to Eq.~(19) of Andreev and
Lifshitz. \cite{andreev69}

In the particular case in which the lattice sites are fixed (so that
${\bf u}=0$) we recover from Eq.~(\ref{eqn.motion.defect}) the
fourth sound modes obtained by Andreev and Lifshitz,
\cite{andreev69} which have the dispersion relation
\begin{equation}
\omega^2 
= {\rs_0}_{ij} \frac{\p \mu}{\p \rho}\bigg|_{w_{ij}} q_i q_j,
\label{fourth.sound}
\end{equation}
where we have used $(\p / \p \rD)_{w_{ij}} = (\p / \p
\rho)_{w_{ij}}$. On the other hand, when there are no defect
fluctuations ($\d \rD = 0$), Eqs.~(\ref{eqn.motion.u}) and
(\ref{eqn.motion.defect}) are combined into
\begin{eqnarray}
\rho_0 \p_t^2 u_i &=&
\frac{\p \lambda_{ji}}{\p w_{lk}}\bigg|_\rho \p_j w_{lk}
- \rho_0 \frac{\p \mu}{\p w_{lk}}\bigg|_\rD \p_i w_{lk}
\nonumber
\\
&&
- \rho_0 \frac{\p \mu}{\p w_{ij}}\bigg|_\rho \p_j w_{kk}.
\end{eqnarray}
A mode analysis of this equation would produce six sound modes of an
anisotropic normal solid. We see that without defects there are no
additional sound modes, as expected.

\section{Density-Density Correlation Function of a Model Supersolid}

In this section we will calculate the density-density correlation
function of a model supersolid, a measurable quantity in a light
scattering experiment. However, before delving into the calculations
for a supersolid let's first review what's revealed by scattering
light from a structureless, normal fluid [for example, see
Ref.~\onlinecite{fleury69}]. The mode counting for the fluid is
simple--there are five collective modes, due to conservation of
mass, energy, and three components of momentum (in three
dimensions). The five collective modes are a pair of transverse
momentum diffusion modes and three longitudinal modes: a pair of
propagating sound modes and a thermal diffusion mode. The density
fluctuations important for light scattering only couple to the
longitudinal modes, so three modes are observed: the diffusion mode
appears as the Rayleigh peak $\omega = 0$ and the pair of sound
modes as the Brillouin doublet at $\omega = \pm c q$ (with a sound
speed $c$). In the absence of dissipation these peaks are
$\d$-functions; dissipation turns each $\d$-function into a
Lorentzian of width $D q^2$, with $D$ being an attenuation
coefficient.

What happens in a superfluid? In addition to the five conserved
densities that exist in a normal fluid there is a broken gauge
symmetry, so from mode counting we conclude there are six collective
modes. Two of these are transverse momentum diffusion modes (just as
for the normal fluid), leaving \emph{four} longitudinal modes for
the superfluid: a pair of propagating first sound modes, and a new
pair of propagating second sound modes. In essence, the central
Rayleigh peak in the normal fluid splits into a new Brillouin
doublet upon passing into the superfluid phase. This remarkable
phenomenon has been observed in light scattering experiments on
$^4$He. \cite{winterling73, tarvin77} We show below that an
analogous splitting occurs in a supersolid, and should be observable
in a light scattering experiment.

\subsection{Dynamics of supersolid without dissipation}

To facilitate the calculation of the density-density correlation
function for a supersolid we'll make two simplifying assumptions:
the solid is isotropic, and two dimensional. The isotropy causes the
transverse and longitudinal modes to neatly decouple; the two
dimensionality results in only one pair of propagating transverse
modes, rather than two pair. Since we're interested in longitudinal
fluctuations, the latter simplification is of little consequence to
the main results of this section. With these assumptions, the
thermodynamic relations are
\begin{equation}
\frac{\p \lambda_{ji}}{\p w_{lk}}\bigg|_\rho = \llam \d_{ji} \d_{lk}
+ \mmu(\d_{il}\d_{jk} + \d_{ik}\d_{jl}),
\label{thermo.relation.1}
\end{equation}
\begin{equation}
\frac{\p \mu}{\p w_{ij}}\bigg|_\rho = \gamma \d_{ij},
\label{thermo.relation.2}
\end{equation}
\begin{equation}
\frac{\p \mu}{\p \rD}\bigg|_{w_{ij}} = \frac{\p \mu}{\p \rho}\bigg|_{w_{ij}} =
\frac{1}{\rho_0^2 \chi},
\label{thermo.relation.3}
\end{equation}
where $\chi$ is the isothermal compressibility at constant strain,
$\gamma$ is a phenomenological coupling constant between the strain
and the density, and $\llam$ and $\mmu$ are the bare Lam\'e
coefficients at constant density. Then in Fourier space the
Lagrangian density, Eq.~(\ref{SS.quad.lag.defect}), reduces to
\begin{widetext}
\begin{equation}
\lag_\text{SS} =
\half \left( \begin{array}{ccc}
\d \rD({\bf Q}) &
\theta({\bf Q}) &
u_L({\bf Q})
\end{array} \right)
{\bf A}
\left( \begin{array}{c}
\d \rD(-{\bf Q}) \\
\theta(-{\bf Q}) \\
u_L(-{\bf Q})
\end{array} \right)
+ \half \left( {\rho_n}_0 \omega_n^2 + \mmu q^2 \right) u_T({\bf Q})
u_T(-{\bf Q}), \label{SS.quad.lag.Fourier}
\end{equation}
where $\omega_n = i\omega$, ${\bf Q} = ({\bf q}, \omega_n)$, $u_L =
({\bf q} \cdot {\bf u}) / q$ with $q = |{\bf q}|$, ${\bf u}_T = {\bf
u} - (u_L / q) {\bf q}$, and
\begin{equation}
{\bf A} = \left( \begin{array}{ccc}
\frac{1}{\rho_0^2\chi} & -\omega_n &
-i q \left(\gamma - \frac{1}{\rho_0\chi}\right) \\
\omega_n & q^2{\rs}_0 & i \omega_n q {\rs}_0 \\
i q \left(\gamma - \frac{1}{\rho_0\chi}\right) &
i \omega_n q {\rs}_0 & \rn_0 \omega_n^2 + q^2 \left(\lambda +
\frac{1}{\chi} - 2 \rho_o \gamma\right)
\end{array} \right),
\label{matrix}
\end{equation}
\end{widetext}
where $\lambda \equiv \llam + 2 \mmu$. The collective modes are
determined from the determinant $\Delta_{\bf A}$ of ${\bf A}$:
\begin{eqnarray}
\Delta_{\bf A} &=& \rn_0 \omega_n^4
+ \left[ \lambda + \rn_0 \left( \frac{1}{\rho_0\chi} - 2 \gamma\right)\right] q^2 \omega_n^2
\nonumber
\\
&&
- \rs_0 \left(\gamma^2 - \frac{\lambda}{\chi\rho_0^2}\right)q^4.
\end{eqnarray}
Setting $\Delta_{\bf A}=0$, we find the longitudinal first sound
speed $c_L$ and second sound speed $c_2$:
\begin{equation}
\begin{split}
c_L^2 &=
\frac{\lambda}{2\rn_0} + \frac{1}{2\rho_0\chi} - \gamma
\\
& + \half \sqrt{ \left(\frac{\lambda}{\rn_0} + \frac{1}{\rho_0\chi}
- 2 \gamma \right)^2 -\frac{4\rs_0}{\rn_0}
\left[\frac{\lambda}{\chi\rho_0^2}-\gamma^2\right]}\ ,
\label{long.sound.vel}
\end{split}
\end{equation}
\begin{equation}
\begin{split}
c_2^2 &=
\frac{\lambda}{2\rn_0} + \frac{1}{2\rho_0\chi} - \gamma
\\
& - \half \sqrt{ \left(\frac{\lambda}{\rn_0} + \frac{1}{\rho_0\chi}
- 2 \gamma \right)^2 -\frac{4 \rs_0}{\rn_0}
\left[\frac{\lambda}{\chi\rho_0^2}-\gamma^2\right]}\ .
\label{second.sound.vel}
\end{split}
\end{equation}
In particular, when $\rs_0 = 0$ (normal solids), $c_2$ vanishes, and we only have
\begin{equation}
c_\text{NS}^2 = (\llam + 2 \mmu + 1 / \chi) / \rho_0 - 2 \gamma,
\label{NS.sound.speed}
\end{equation}
which agrees with the longitudinal sound speed obtained by Zippelius
\emph{et al.} \cite{zippelius80} once we identify $\llam =
\lambda^\text{ZHN} + 2 \gamma^\text{ZHN} + 1 / \chi$ and $\gamma =
(\gamma^\text{ZHN} + 1 / \chi) / \rho_0$. Moreover, when the Lam\'e
coefficients and the coupling constant $\gamma$ vanish we recover
the sound speed of a normal fluid. As discussed earlier, there is
one pair of transverse sound modes with speed
\begin{equation}
c_T = \sqrt{\frac{\mmu}{\rn_0}}.
\end{equation}

Finally, we can calculate the correlation functions from
Eq.~(\ref{SS.quad.lag.Fourier}):
\begin{equation}
\left<\d \rD({\bf Q}) \d \rD(-{\bf Q})\right> 
= \rs_0 q^2
\frac{\rho_0 \omega_n^2 + (\lambda -2\rho_0\gamma + 1/\chi)q^2}
{\Delta_{\bf A}},
\label{defect.defect.correlation.fcn}
\end{equation}
\begin{equation}
\left<\d \rD({\bf Q}) u_L(-{\bf Q})\right> =
iq\frac{\rs_0}{\rho_0}
\frac{\rho_0 \omega_n^2 - (\rho_0\gamma-1/\chi)q^2}{\Delta_{\bf A}},
\label{defect.uL.correlation.fcn}
\end{equation}
and
\begin{equation}
\left<u_L({\bf Q}) u_L(-{\bf Q})\right> =
\frac{1}{\rho_0^2\chi}
\frac{\rho_0^2\chi\omega_n^2 + \rs_0 q^2}{\Delta_{\bf A}}.
\label{uL.uL.correlation.fcn}
\end{equation}
Since the density fluctuation is related to the defect density fluctuation and
the strain tensor by Eq.~(\ref{quadrho1}), the density-density correlation function becomes
\begin{eqnarray}
\left<\d \rho({\bf Q}) \d \rho(-{\bf Q})\right> &=&
A \left( \frac{1}{i \omega - c_Lq} - \frac{1}{i \omega + c_Lq} \right)
\nonumber
\\
&+&
B \left( \frac{1}{i \omega - c_2q} - \frac{1}{i \omega + c_2q} \right),
\label{correlation.fcn}
\end{eqnarray}
where
\begin{equation}
A = - q \frac{\rho_0 \rn_0 c_L^2 - \rs_0 \lambda}
{2c_L \rn_0 (c_L^2 - c_2^2)},
\end{equation}
\begin{equation}
B = - q \frac{\rho_0 \rn_0 c_2^2 - \rs_0 \lambda}
{2c_2 \rn_0 (c_2^2 - c_L^2)}.
\end{equation}
Then, by performing the analytic continuation $i \omega_n = \omega +
i \delta$, the density-density response function can be obtained
from the imaginary part of the density-density correlation function:
\begin{eqnarray}
\chi^{\prime\prime}_{\rho\rho}({\bf q}, \omega) &=&
- \pi A \bigg[ \d(\omega - c_Lq) - \d(\omega + c_Lq) \bigg]
\nonumber
\\
&&
- \pi B \bigg[ \d(\omega - c_2q) - \d(\omega + c_2q) \bigg],
\label{non.dissipative.susceptibility}
\end{eqnarray}
where we have used the identity
\begin{equation}
\frac{1}{\omega^\prime-\omega-i\epsilon}=
P \frac{1}{\omega^\prime-\omega} +i\pi \delta \left( \omega-\omega^\prime \right).
\end{equation}
It is easy to show that the response function satisfies the
thermodynamic sum rule (for the derivation of the static correlation
function see Appendix~B)
\begin{equation}
\int^\infty_{-\infty} \frac{d\omega}{\pi} \frac{\chi^{\prime
\prime}_{\rho\rho}({\bf q},\omega)}{\omega} =
\frac{\rho_0^2\chi\lambda}{\lambda - \rho_0^2\gamma^2 \chi},
\label{thermodynamic_sum_rule}
\end{equation}
and the f-sum rule
\begin{equation}
\int^\infty_{-\infty} \frac{d\omega}{\pi} \omega \chi^{\prime
\prime}_{\rho\rho}({\bf q},\omega) = \rho_0 q^2. \label{fsum_rule}
\end{equation}

\subsection{Dynamics of supersolid with dissipation}

We continue our discussion of the density correlation and response
functions by including dissipative terms in the equations of motion.
As mentioned above, the dissipative terms will broaden the
$\delta$-function peaks in the density response function. In
addition, as noted by Martin et al.\cite{martin72}, the dissipation
is necessary to identify the ``missing'' defect diffusion mode in
normal solids. The dissipative hydrodynamic equations of motion for
a supersolid were first obtained by Andreev and
Lifshitz,\cite{andreev69} who used standard entropy-production
arguments to generate the dissipative terms. For an isotropic
supersolid (including the nonlinear term neglected by Andreev and
Lifshitz) we have (with the new dissipative terms on the right hand
side)
\begin{equation}
\p_t \rho + \p_i j_i = 0,
\label{eqn.motion.dissipation1}
\end{equation}
\begin{eqnarray}
\p_t j_i + \partial_j \Pi_{ij} &=&
 \zeta \p_i \p_k \vn_k + \eta \p^2 \vn_i 
\nonumber
\\
&&
- \Sigma \p_i \p_k \bigg[\rs (\vn_k - \vs_k) \bigg],
\end{eqnarray}
\begin{equation}
\p_t u_i - \vn_i + \vn_k \p_k u_i + u_i \p_k \vn_k = \Gamma \p_k
\lambda_{ki},
\end{equation}
\begin{eqnarray}
\p_t \vs_i + \p_i \left( \mu + \half \vs^2 \right) &=&
- \Lambda \p_i \p_k \bigg[ \rs (\vn_k - \vs_k) \bigg] 
\nonumber
\\
&&
+ \Sigma \p_i \p_k \vn_k ,
\label{eqn.motion.dissipation4}
\end{eqnarray}
where ${\bf j} = \rn \bvn + \rs \bvs$ is the total mass current,
$\Sigma$ and $\Lambda$ coefficients of viscosity, $\zeta$ the bulk
viscosity coefficient, $\eta$ the shear viscosity coefficient,
$\Gamma$ the diffusion coefficient for defects.

We next linearize the dissipative hydrodynamic equations by
considering small fluctuations from the equilibrium values. Writing
$\d \mu$ and $\lambda_{ij}$ in terms of $\d \rho$ and $\d w_{ij}$,
\begin{equation}
\d \mu = \frac{1}{\rho_0^2 \chi} \d \rho + \gamma w_{ii},
\end{equation}
\begin{equation}
\d \lambda_{ij} = \gamma \d_{ij} \d \rho + \llam \d_{ij} w_{kk}
+ \mmu (w_{ij} + w_{ji}).
\end{equation}
We replace $\d \mu$ and $\d \lambda_{ij}$ into
Eqs.~(\ref{eqn.motion.dissipation1})-(\ref{eqn.motion.dissipation4}),
and divide them into transverse and longitudinal parts. The
equations of motion for transverse parts are
\begin{equation}
\rn_0 \p_t \vn^T - \mmu \p^2 u^T - \eta \p^2 \vn^T = 0,
\label{trans.eqn.motion.dissipation1}
\end{equation}
where $\p \equiv \sqrt{\p_i^2}$, and
\begin{equation}
\p_t u^T - \vn^T - \mmu \Gamma \p^2 u^T = 0.
\label{trans.eqn.motion.dissipation2}
\end{equation}
These equations support a propagating transverse sound mode with the
transverse sound speed $c_T = \sqrt{\mmu/\rn_0}$, as obtained in the
previous section, and an attenuation constant $\Gamma_T = \eta +
\rn_0 \mmu \Gamma$. Next, the longitudinal equations of motion are
\begin{equation}
\p_t \d \rho + \rs_0 \p \vs^L + \rn_0 \p \vn^L = 0,
\label{long.eqn.motion.dissipation1}
\end{equation}
\begin{equation}
\begin{split}
\rn_0 \p_t \vn^L &+ \left( \frac{\rn_0}{\rho_0^2 \chi} - \gamma \right) \d \rho
- \bigg( \lambda - \rn_0 \gamma \bigg) \p^2 u^L
\\
- \bigg( \tilde{\zeta} - &2 \rs_0 \sigma - \rs_0^2 \Lambda \bigg) \p^2 \vn^L
- \rs_0 \sigma \p^2 \vs^L = 0,
\end{split}
\label{long.eqn.motion.dissipation2}
\end{equation}
\begin{equation}
\p_t u^L - \vn^L - \gamma \Gamma \p \d \rho - \lambda \Gamma \p^2 u^L
= 0,
\label{long.eqn.motion.dissipation3}
\end{equation}
\begin{equation}
\p_t \vs^L + \frac{1}{\rho_0^2 \chi} \p \d \rho + \gamma \p^2 u^L
- \sigma \p^2 \vn^L
- \rs_0 \Lambda \p^2 \vs^L = 0,
\label{long.eqn.motion.dissipation4}
\end{equation}
where $\sigma \equiv \Sigma - \rs_0 \Lambda$,
and $\tilde{\zeta} \equiv \zeta + \eta$. The Laplace-Fourier transform of
Eqs.~(\ref{long.eqn.motion.dissipation1}) - (\ref{long.eqn.motion.dissipation4})
yields
\begin{equation}
{\bf C}
\left( \begin{array}{c}
\d \rho ({\bf q}, z) \\
\vn^L ({\bf q}, z) \\
u^L ({\bf q}, z) \\
\vs^L ({\bf q}, z)
\end{array} \right) =
\left( \begin{array}{c}
\d \rho ({\bf q}) \\
\vn^L ({\bf q}) \\
u^L ({\bf q}) \\
\vs^L ({\bf q}),
\end{array} \right)
\end{equation}
where
\begin{widetext}
\begin{equation}
{\bf C} =
\left( \begin{array}{cccc}
-iz & iq\rn_0 & 0 & iq\rs \\
iq\left( \frac{1}{\rho_0^2\chi} - \frac{\gamma}{\rn_0} \right) &
-iz + q^2 \frac{1}{\rn_0} \left( \tilde{\zeta} - 2 \rs_0 \sigma - \rs_0^2
\Lambda \right) &
q^2 \frac{1}{\rn_0} \left( \lambda - \rn_0 \gamma \right) &
q^2 \frac{\rs_0}{\rn_0} \sigma \\
-iq\gamma \Gamma & -1 & -iz + q^2 \lambda \Gamma & 0 \\
iq\frac{1}{\rho_0^2 \chi} & q^2 \sigma & -q^2 \gamma & -iz + q^2 \rs_0 \Lambda
\end{array} \right).
\label{C}
\end{equation}
\end{widetext}
From Eq.~(\ref{C}) we calculate two sound speeds $c_L$,
Eq.~(\ref{long.sound.vel}), and $c_2$, Eq.~(\ref{second.sound.vel}),
with two attenuation constants,
\begin{equation}
D_L = -\frac{1}{\rn_0(c_L^2 - c_2^2)} \left( c_L^2 n_1 + n_2 \right),
\end{equation}
\begin{equation}
D_2 = \frac{1}{\rn_0(c_L^2 - c_2^2)} \left( c_2^2 n_1 + n_2 \right),
\end{equation}
where
\begin{equation}
n_1 \equiv \tilde{\zeta} - 2\rs_0\sigma + \rn_0\Gamma\lambda
+ \rs_0(\rn_0 - \rs_0)\Lambda,
\end{equation}
\begin{equation}
\begin{split}
n_2 \equiv& \frac{1}{\rho_0^2\chi} \Bigg\{
2 \rho_0\rs_0 (\rho_0\chi\gamma - 1)\sigma
+ \rho_0 \rn_0(\lambda - \rho_0^2\chi\gamma^2)\Gamma
\\
&
+\rs_0 \tilde{\zeta}
+ \rho_0 \rs_0 \bigg[ \rn_0 - \rs_0 + \rho_0\chi (\lambda - 2\gamma\rn_0)
\bigg] \Lambda
\Bigg\}.
\end{split}
\end{equation}
Now we can see that when $\rs=0$ (a normal solid), the second sound
modes disappear but there remains the defect diffusion mode with the
diffusion constant $D_2 = (\lambda - \rho_0^2 \chi
\gamma^2)\Gamma/\rho_0 \chi c_L^2$.

We also calculate the density-density Kubo function from
Eq.~(\ref{C}), \cite{forster75}
\begin{eqnarray}
K_{\rho \rho} ({\bf q},z) &=& \frac{\chi_{\rho \rho}({\bf q})}{k_B T}
\frac{i z^3 + b_{\rho \rho} z^2 + d_{\rho \rho} q^2 z
+ e_{\rho \rho} q^2}{Z}
\nonumber
\\
&&
+ \frac{\chi_{u_L \rho}({\bf q})}{k_B T}
\frac{d_{\rho u_L} q^2 z
+ e_{\rho u_L} q^2}{Z},
\label{Kubo.function}
\end{eqnarray}
where the static susceptibilities $\chi_{\rho \rho}$ and $\chi_{u_L
\rho}$ in Eq.~(\ref{Kubo.function}) are given in Appendix~B, and
\begin{equation}
Z \equiv
(z^2 - c_L^2 q^2 + i z q^2 D_L)(z^2 - c_2^2 q^2 + i z q^2 D_2),
\end{equation}
\begin{equation}
b_{\rho \rho} = - \Gamma \gamma q^2 - \frac{\rs_0(\rn_0-\rs_0)}{\rn_0}
 \Lambda q^2
- \frac{\tilde{\zeta} - 2\rs_0 \sigma}{\rn_0} q^2,
\label{b.rho.rho}
\end{equation}
\begin{equation}
d_{\rho \rho} = -i \frac{\lambda}{\rn_0} + i \gamma,
\label{d.rho.rho}
\end{equation}
\begin{equation}
e_{\rho \rho} = \frac{\rs_0 \Lambda \lambda}{\rn_0} q^2
- \rs_0 \Lambda \gamma q^2
+ \frac{\rs_0 \sigma \gamma}{\rn_0} q^2,
\label{e.rho.rho}
\end{equation}
\begin{equation}
d_{\rho u_L} = (\lambda - \rho_0 \gamma) q,
\label{d.rho.uL}
\end{equation}
\begin{eqnarray}
e_{\rho u_L} &=&
i \rs_0 \left( \Lambda - \frac{\sigma}{\rn_0} \right) \lambda q^3
\nonumber
\\
&&
+ i \frac{\rs_0}{\rn_0} \bigg[ 2\sigma \rho_0 - \tilde{\zeta}
- \rho_0 \Lambda (\rho_0 - 2\rs_0) \bigg] \gamma q^3.
\label{e.rho.uL}
\end{eqnarray}
Then the susceptibility $\chi_{\rho \rho}^{\prime \prime} ({\bf
q},\omega)$ can be obtained from the real part of
Eq.~(\ref{Kubo.function}), \cite{chaikin95, forster75}
\begin{equation}
\begin{split}
&\frac{\chi_{\rho \rho}^{\prime \prime} ({\bf q},\omega)}{\omega} =
- \frac{iq^4 c_L^2 D_L I_1(q)}
{(\omega^2 - c_L^2 q^2)^2 + (\omega q^2 D_L)^2}
\\
&
- \frac{iq^4 c_2^2 D_2 I_2(q)}
{(\omega^2 - c_2^2 q^2)^2 + (\omega q^2 D_2)^2}
+ \frac{(\omega^2 - c_L^2 q^2) I_3(q)}
{(\omega^2 - c_L^2 q^2)^2 + (\omega q^2 D_L)^2}
\\
&
+ \frac{(\omega^2 - c_2^2 q^2) I_4(q)}
{(\omega^2 - c_2^2 q^2)^2 + (\omega q^2 D_2)^2},
\label{susceptibility}
\end{split}
\end{equation}
where $I_1(q)$, $I_2(q)$, $I_3(q)$, and $I_4(q)$ are given in
Appendix~C.

 Equation (\ref{susceptibility}) is one of the central results of this paper,
it's worth exploring some of its features and limits. First, one can
show that Eq.~(\ref{susceptibility}) satisfies both the
thermodynamic sum rule, Eq.~(\ref{thermodynamic_sum_rule}), and
f-sum rule, Eq.~(\ref{fsum_rule}). Second, the first and second
terms in Eq.~(\ref{susceptibility}) produce two Brillouin doublets
centered at $\omega = \pm c_L q$ and $\omega = \pm c_2 q$ with
widths $D_L q^2$ and $D_2 q^2$, respectively. The third and fourth
terms in Eq.~(\ref{susceptibility}) are negligible near the
Brillouin doublets, and in fact these terms vanish in the limit of
zero dissipation. Therefore the non-dissipative density-density
correlation function, Eq.~(\ref{non.dissipative.susceptibility}), is
obtained from the first two terms in the limit $D_L, D_2 \rightarrow
0$. Finally, for normal solids ($\rs =0$), the second term in
Eq.~(\ref{susceptibility}) vanishes, and there is only one Brillouin
doublet due to the longitudinal first sound modes. In this case the
fourth term in Eq.~(\ref{susceptibility}) becomes the Rayleigh peak
of the defect diffusion mode centered at $\omega = 0$. Therefore, we
see that in analogy with a superfluid,\cite{winterling73} \emph{the
defect diffusion peak that exists in a normal solid will split into
a Brillouin doublet of second sound modes upon entering the
supersolid phase}.

\begin{figure}[tp]
\centering
\includegraphics[totalheight=7cm,width=7cm]{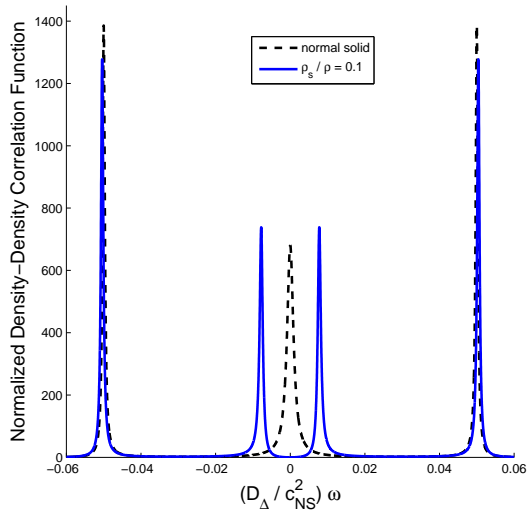}
\caption{\label{figure1} (Color online). Density-density correlation
functions of a normal solid (black dashed line) and a supersolid
(blue solid line). The supersolid fraction is assumed to be 10 \%.}
\end{figure}

\begin{figure}[tp]
\centering
\includegraphics[totalheight=7cm,width=7cm]{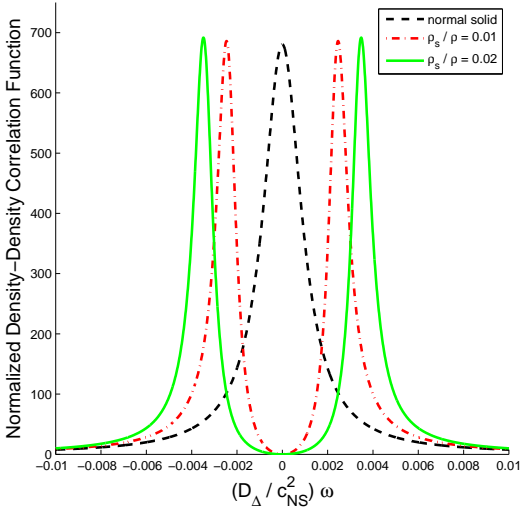}
\caption{\label{figure2} (Color online). Splitting of the Rayleigh
peak (black dashed line)due to the defect diffusion mode in the
normal solid phase into the Brillouin doublet of the second sound
modes in the supersolid phase. The red dash-dot line is for
$\rs/\rho$ = 1\%, and the green solid line $\rs/\rho$ = 2\%.}
\end{figure}

To get a sense of the size of this effect, let's substitute some
physically realistic numbers into the correlation function. Assuming
$\rs\ll\rho_0$ and $\gamma = \Lambda = \Sigma = 0$, we have
\begin{eqnarray}
I_1(q) &=& -I_2(q) + i \alpha \frac{\rho_0}{c_\text{NS}^2}
\nonumber
\\
&=& i \frac{\rho_0}{c_\text{NS}^2} - 2 i \frac{(\alpha -
1)^2}{\alpha^2} \frac{\rs_0}{c_\text{NS}^2} + {\cal
O}\left(\frac{\rs_0^2}{\rho_0^2}\right),
\end{eqnarray}
\begin{eqnarray}
I_3(q) &=& -I_4(q)
\nonumber
\\
&=&
- \frac{(\alpha - 1)^2}{\alpha} \frac{\rho_0}{c_\text{NS}^2} D_\Delta q^2
\nonumber
\\
&&
+2 q^2  \frac{\alpha -1}{\alpha^2} \bigg[
\frac{(\alpha -1)^2(\alpha -3)}{\alpha} D_\Delta + D_l
\bigg] \frac{\rs_0}{c_\text{NS}^2}
\nonumber
\\
&& + {\cal O}\left(\frac{\rs_0^2}{\rho_0^2}\right),
\end{eqnarray}
where the longitudinal sound speed of normal solid $c_\text{NS}$ is
given in Eq.~(\ref{NS.sound.speed}), the defect diffusion constant
$D_\Delta \equiv \Gamma / \chi$, and $\alpha \equiv \rho_0 \chi
c_\text{NS}^2$. We show in Fig.~\ref{figure1} the normalized
density-density correlation functions of a normal solid and a
supersolid. We have used the first sound speed $c_\text{NS} = 550$
m/s, the density $\rho = 0.19048$ g/cm$^3$, the isothermal
compressibility $\chi = 0.29615\times10^{-8}$ cm s$^2$/g  for $^4$He
solid\cite{wanner70, trickey72} at the molar volume 21 cm$^3$/mole,
the viscosity of $^4$He \emph{fluid} of $2\times10^{-5}$ g/cm s, the
typical wave number involved in light scattering $q^{-1} = 100$ nm,
and $\Gamma = 8\times10^{-11}$ cm$^3$s/g.
%
%
In Fig.~\ref{figure2} we show the splitting of the Rayleigh peak due
to defect diffusion in a normal solid into a Brillouin doublet of
second sound modes, for two values of the supersolid fraction. 

\section{Conclusion}

Starting from general conservation laws and symmetry principles we
derived the effective action for a supersolid---a state of matter
with simultaneous broken translational symmetry and Bose
condensation. The resulting effective action in Eq.~(\ref{lag4}) is
one of the two important results of this work, and will be further
developed in subsequent work on vortex and dislocation dynamics in
supersolids. In this work, however, we used the linearized version
of this action to calculate the second of our important results--the
density-density correlation and response functions for a model
supersolid (with isotropic elastic properties), see Eq.
(\ref{susceptibility}). In complete analogy with a superfluid, we
showed that the onset of supersolidity causes the zero-frequency
defect diffusion mode to split into propagating second sound modes,
and from our calculation we can determine the spectral weight in
these modes as well as the weight in the ``normal'' longitudinal
sound waves in a solid.

\begin{acknowledgments}
The authors would like to thank K. Dasbiswas, J. Dufty, P. Goldbart, D.
Goswami, Y. Lee, M. Meisel, and J. Toner for helpful discussions and
comments. A.T.D. would like to thank the Aspen Center for Physics,
where part of this work was completed. This work was supported by
NSF Grant No. DMR-0705690.
\end{acknowledgments}

\appendix

\section{Variational principle for an ideal fluid and the Gross-Pitaevskii action}

This appendix demonstrates the simplicity of the variational
principle for deriving the hydrodynamic equations of motion and the
Lagrangian density for continuum systems. Consider the simplest case
of an ideal fluid (IF) which is irrotational, inviscid and
incompressible. The Lagrangian density for the ideal fluid is
\begin{equation}
\lag_\text{IF} = \half \rho v^2 - U_\text{IF}(\rho),
\label{ideal.fluid.lagrangian}
\end{equation}
where $\rho$ is the mass density, ${\bf v}$ the velocity field, and
$U_\text{IF}$ the internal energy density which satisfies $d
U_\text{IF} = \mu d \rho$. From the Lagrangian density we construct
the action $S = \int dt\int d^3x\, \lag_\text{IF}$. The variational
principle states that the equations of motion are derived from
variations of the action with respect to all the dynamical
variables. The naive application of this principle to the ideal
fluid leads to the trivial equation of motion ${\bf v} = 0$. The
difficulty is that the dynamical variables $\rho$ and ${\bf v}$ are
not independent, but constrained by the conservation of mass,
\begin{equation}
\p_t \rho + \p_i (\rho v_i) =0.
\label{IF.continuity}
\end{equation}
This \emph{constraint} is incorporated into the Lagrangian density
by introducing a Lagrange multiplier $\phi$:
\begin{equation}
\lag_\text{IF} = \half \rho v^2 - U_\text{IF}(\rho)
+ \phi \bigg[ \p_t \rho + \p_i (\rho v_i) \bigg].
\label{ideal.fluid.lagrangian1}
\end{equation}
Then the equations of motion are obtained from variations of the
action $S[\rho, {\bf v},\phi]$ with respect to $\rho$, ${\bf v}$,
and $\phi$:
\begin{equation}
\frac{\d S}{\d \rho} =
\half v^2 -\frac{\p U_\text{IF}}{\p \rho} - \frac{D \phi}{D t} = 0,
\label{IF.eqn.phi}
\end{equation}
\begin{equation}
\frac{\d S}{\d v_i} = \rho v_i - \rho \p_i \phi = 0,
\label{IF.eqn.vel}
\end{equation}
\begin{equation}
\frac{\d S}{\d \phi} =  \p_t \rho + \p_i (\rho v_i) = 0.
\end{equation}
From Eq.~(\ref{IF.eqn.vel}) we obtain the velocity field
\begin{equation}
v_i = \p_i \phi,
\label{IF.vel1}
\end{equation}
which implies that there is no vorticity, as expected for an ideal
fluid. We can derive the Euler equation from Eq.~(\ref{IF.eqn.phi})
by taking its gradient, using Eq.~(\ref{IF.vel1}), and then the
Gibbs-Duhem relation to obtain
\begin{equation}
\rho \frac{D v_i}{D t} = - \p_i P,
\end{equation}
where $P$ is the pressure. The Lagrangian density may be cast into
an equivalent form by substituting ${\bf v} = \nabla \phi$ into
Eq.~(\ref{ideal.fluid.lagrangian1}) and integrating by parts, with
the result
\begin{equation}
\lag_\text{IF} = -\rho \p_t \phi - \half \rho (\p_i \phi)^2 -
U_\text{IF}(\rho). \label{IF.lag3}
\end{equation}

For comparison, consider a system of weakly interacting bosons with
a condensate wave function $\psi({\bf r}, t)$ that satisfies
Gross-Pitaevskii equation \cite{dalfovo99}
\begin{equation}
i\hbar \frac{\partial \psi}{\partial t} = - \frac{\hbar^2}{2m}
\nabla^2 \psi  + g |\psi|^2 \psi.
\end{equation}
This equation of motion can be derived from the Lagrangian density
\begin{equation}
\lag_\text{GP} = \frac{i\hbar}{2} \left[ \psi^* \p_t \psi - \psi
\p_t \psi^* \right]
-\frac{\hbar^2}{2m}(\nabla\psi^*)\cdot(\nabla\psi)
 - \frac{g}{2} (\psi^* \psi)^2.
\label{GPlagSF}
\end{equation}
Taking $\psi = \sqrt{n} e^{i \tilde{\phi}}$ with the number density
$n$, the Gross-Pitaevskii Lagrangian density becomes
\begin{equation}
\lag_\text{GP} = \frac{i\hbar}{2} \p_t n - \hbar n \p_t \tilde{\phi}
-\frac{\hbar^2}{8mn} (\p_i n)^2 - \frac{\hbar^2}{2m}n (\p_i
\tilde{\phi})^2 - \frac{g}{2} n^2. \label{GPlagSF1}
\end{equation}
The first term contributes $i\hbar N/2$ to the action (with $N$ the
number of particles) and does not contribute to the dynamics.
Identifying $\rho = m n$ and $\phi = (\hbar / m) \tilde{\phi}$, we
see that the Gross-Pitaevskii Lagrangian density,
Eq.~(\ref{GPlagSF1}), is identical to the ideal fluid Lagrangian
density, Eq.~(\ref{IF.lag3}), with
\begin{equation}
U_\text{IF}(\rho) = \frac{\hbar^2}{2m^2}(\nabla \sqrt{\rho})^2 +
\frac{g}{2m^2} \rho^2.
\end{equation}

\section{Thermodynamic Relations and the Static Correlation Functions of Supersolids}

In this appendix we calculate the thermodynamic relation for the
potential energy density in Eq.~(\ref{lagrangian1}). Given the
Lagrangian density, Eq.~(\ref{lagrangian1}), the total energy
density for a supersolid is defined as the sum of the kinetic energy
densities and the internal energy density
\begin{eqnarray}
E_\text{SS} &=& \half \rs_{ij} \vs_i \vs_j
+ \half (\rho \d_{ij} - \rs_{ij} ) \vn_i\vn_j
\nonumber
\\
&&
+ U_\text{SS}(\rho, \rs_{ij}, s, R_{ij}).
\label{SStotalenergy1}
\end{eqnarray}
Following Andreev and Lifshitz \cite{andreev69}, this total energy
density can be related to the energy density $\epsilon$ measured in
the frame where the super-component is at rest as
\begin{equation}
E_\text{SS} = \half \rho v_s^2 + (\rho \d_{ij} - \rs_{ij}) (\vn_j - \vs_j) \vs_i
+ \epsilon,
\label{SStotalenergy2}
\end{equation}
where $\epsilon$ has a thermodynamic relation
\begin{eqnarray}
d \epsilon &=&
T ds + \mu d \rho -\lambda_{ik} d R_{ik}
\nonumber
\\
&&
+ (\vn_i - \vs_i) d \left[(\rho \d_{ij} - \rs_{ij})(\vn_j - \vs_j)\right].
\label{SS.energy.1}
\end{eqnarray}
We can obtain the thermodynamic relation for the total energy
$E_\text{SS}$ by differentiating Eq.~(\ref{SStotalenergy2}) and
using Eq.~(\ref{SS.energy.1}) for $d\epsilon$, with the result
\begin{eqnarray}
dE_\text{SS} &=& T ds - \lambda_{ik} d R_{ik}
- (\vn_i - \vs_i) \vn_j d\rs_{ij}
\nonumber
\\
&&
+ \left[ \mu + \half (2 \vn_i^2 -2 \vn_i \vs_i + \vs_i^2 ) \right] d \rho
\nonumber
\\
&&
+ \rs_{ij} \vs_j d \vs_i
+ (\rho \d_{ij} - \rs_{ij}) \vn_i d \vn_j.
\label{SS.thermo.2}
\end{eqnarray}
This thermodynamic relation agrees with Eq.~(2.18) of Saslow
\cite{saslow77} and Eq.~(2.1) of Liu \cite{liu78} after identifying
$\epsilon^\text{Saslow, Liu} = E_\text{SS}$, and $\mu^\text{Saslow,
Liu} = \mu - \vn_i \vs_i + \vs_i^2 / 2$. Then the differentiation of
Eq.~(\ref{SStotalenergy1}) and the use of Eq.~(\ref{SS.thermo.2})
give the thermodynamic relation for $U_\text{SS}$,
Eq.~(\ref{thermo.Uss}).

For a supersolid at rest we can expand the free energy $F_\text{SS}
= E_\text{SS} - TS$ up to the second order in the density
fluctuations $\d \rho$ and the strains $w_{ij}=\partial_i u_j$:
\begin{equation}
F_\text{SS} = \half \frac{\p \mu}{\p \rho}\bigg|_{w_{ij}} (\d
\rho)^2 + \frac{\p \mu}{\p w_{ij}}\bigg|_{\rho} \d \rho \, w_{ij} +
\half \frac{\p \lambda_{ij}}{\p w_{lk}}\bigg|_{\rho} w_{ij}w_{lk}.
\label{SS.thermo.4}
\end{equation}
Using Eqs.~(\ref{thermo.relation.1}) - (\ref{thermo.relation.3}) for
an isotropic supersolid, the free energy (in Fourier space) can be
written as
\begin{equation}
F_\text{SS} = \half \mmu q^2 u_T^2
+\half \left( \begin{array}{cc}
\d \rho({\bf q}) &
u_L({\bf q})
\end{array} \right)
{\bf B}
\left( \begin{array}{c}
\d \rho(-{\bf q}) \\
u_L(-{\bf q})
\end{array} \right),
\label{SS.thermo.5}
\end{equation}
where
\begin{equation}
{\bf B} = \left( \begin{array}{cc}
\frac{1}{\rho_0^2\chi} & - i q \gamma \\
i q \gamma & q^2\lambda
\end{array} \right).
\end{equation}
Then the static correlation functions can be easily read off from
Eq.~(\ref{SS.thermo.5}):
\begin{equation}
\chi_{\rho \rho}({\bf q})
= \beta \left<\d \rho({\bf q}) \d \rho(-{\bf q})\right>
= \frac{\rho_0^2 \chi \lambda}{\lambda - \rho_0^2 \gamma^2 \chi},
\end{equation}
\begin{equation}
\chi_{u_L \rho}({\bf q})
= \beta \left<u_L({\bf q}) \d \rho(-{\bf q})\right>
= \frac{i \rho_0^2 \chi \gamma}{q(\lambda - \rho_0^2 \gamma^2 \chi)}.
\end{equation}

\section{Calculation of the density-density correlation function}

Each term in the Kubo function given in Eq.~(\ref{Kubo.function})
can be separated into the first sound part and the second sound part
by performing a partial fraction expansion,
\begin{equation}
\begin{split}
\frac{ a_{jk} z^3 + b_{jk} z^2 +d_{jk} q^2z +q^2e_{jk}}
{(z^2-c_L^2 q^2 +iz D_L q^2)(z^2-c_2^2 q^2 +iz D_2 q^2)}
\\
=\frac{\tilde{A}_{jk}z+\tilde{B}_{jk}}{z^2-c_L^2 q^2 +iz D_L q^2}+
\frac{\tilde{C}_{jk}z+\tilde{D}_{jk}}{z^2-c_2^2 q^2 +iz D_2 q^2},
\end{split}
\end{equation}
where $j,k = (\rho, u_L)$. Then, $\tilde{A}_{jk}$, $\tilde{B}_{jk}$,
$\tilde{C}_{jk}$ and $\tilde{D}_{jk}$ can be written in terms of
$a_{jk}$, $b_{jk}$, $d_{jk}$ and $e_{jk}$ along with the sound
velocities ($c_L$ and $c_2$) and the attenuation coefficients ($D_L$
and $D_2$):
\begin{widetext}
\begin{equation}
\tilde{A}_{jk} =
\frac{a_{jk} \left[ c_L^4 -c_2^2c_L^2+q^2 D_L (D_L c_2^2 -D_2
c_L^2 ) \right]}
{(c_L^2-c_2^2)^2+q^2(D_L-D_2)(D_L c_2^2 -D_2 c_L^2 )}
+\frac{i b_{jk} (c_2^2D_L-D_2 c_L^2)
+d_{jk}(c_L^2-c_2^2)
+i e_{jk} (D_L-D_2)}
{(c_L^2-c_2^2)^2+q^2(D_L-D_2)(D_L c_2^2 -D_2 c_L^2 )},
\end{equation}
\begin{equation}
\tilde{B}_{jk} =
\frac{i a_{jk} c_L^2 q^2 (D_L c_2^2 -D_2 c_L^2 )
+ b_{jk} c_L^2 (c_L^2-c_2^2)}
{(c_L^2-c_2^2)^2+q^2(D_L-D_2)(D_L c_2^2 -D_2 c_L^2 )}
+ \frac{i d_{jk} c_L^2q^2 (D_L-D_2)
+ e_{jk} \left[ c_L^2-c_2^2 +q^2D_L (D_2-D_L) \right]}
{(c_L^2-c_2^2)^2+q^2(D_L-D_2)(D_L c_2^2 -D_2 c_L^2 )}.
\end{equation}
\end{widetext}
The coefficients $\tilde{C}_{jk}$ and $\tilde{D}_{jk}$ are the same
as $\tilde{A}_{jk}$ and $\tilde{B}_{jk}$, respectively, but two
sound velocities and two attenuation coefficients must be
interchanged. Then the functions defined in the density-density
correlation function, Eq.~(\ref{susceptibility}), are given by
\begin{equation}
I_1({\bf q}) =
\chi_{\rho \rho} ({\bf q}) \tilde{A}_{\rho \rho}
+ \chi_{u_L \rho} ({\bf q}) \tilde{A}_{\rho u_L},
\end{equation}
\begin{equation}
I_2({\bf q}) =
\chi_{\rho \rho} ({\bf q}) \tilde{C}_{\rho \rho}
+ \chi_{u_L \rho} ({\bf q}) \tilde{C}_{\rho u_L},
\end{equation}
\begin{equation}
I_3({\bf q}) =
\chi_{\rho \rho} ({\bf q}) \tilde{B}_{\rho \rho}
+ \chi_{u_L \rho} ({\bf q}) \tilde{B}_{\rho u_L}
-iq^2D_L I_1({\bf q}),
\end{equation}
\begin{equation}
I_4({\bf q}) =
\chi_{\rho \rho} ({\bf q}) \tilde{D}_{\rho \rho}
+ \chi_{u_L \rho} ({\bf q}) \tilde{D}_{\rho u_L}
-iq^2D_2 I_2({\bf q}).
\end{equation}

\end{document}